\documentclass[showpacs,preprintnumbers,amsmath,amssymb,floatfix]{revtex4}

\usepackage{color}   
\usepackage{graphicx}
\usepackage{dcolumn}
\usepackage{bm}

\begin{document}
\def\bbox#1{\hbox{\boldmath${#1}$}}
\def\blambda{{\hbox{\boldmath $\lambda$}}}
\def\eeta{{\hbox{\boldmath $\eta$}}}
\def\bxi{{\hbox{\boldmath $\xi$}}}
\def\bzeta{{\hbox{\boldmath $\zeta$}}}

\title{ Potential Models and Lattice Gauge Current-Current
  Correlators}

\author{Cheuk-Yin Wong$^{1,2}$ \& Horace W. Crater$^{3}$}

\affiliation{${}^1$Physics Division, Oak Ridge National Laboratory, 
Oak Ridge, TN\footnote{wongc@ornl.gov} 37831}

\affiliation{${}^2$Department of Physics, University of Tennessee, 
Knoxville, TN 
37996}

\affiliation{${}^3$The University of Tennessee Space Institute, Tullahoma,
TN 37388\footnote{hcrater@utsi.edu}}

\date{\today}

\begin{abstract}

We compare current-current correlators in lattice gauge calculations
with correlators in different potential models, for a pseudoscalar
charmonium in the quark-gluon plasma.  An important ingredient in the
evaluation of the current-current correlator in the potential model is
the basic principle that out of the set of continuum states, only
resonance states and Gamow states with lifetimes of sufficient
magnitudes can propagate as composite objects and can contribute to
the current-current correlator.  When the contributions from the bound
states and continuum states are properly treated, the potential model
current-current correlators obtained with the potential proposed in
Ref. [11] are consistent with the lattice gauge correlators.  The
proposed potential model thus gains support to be a useful tool to
complement lattice gauge calculations for the study of $Q\bar Q$
states at high temperatures.

\end{abstract}

\pacs{ 25.75.-q 25.75.Dw }
                                                                         
\maketitle

\section{Introduction}
                                                                            
Potential models have often been used to describe bound states of
quark-antiquark pairs.  The basic idea is that a short range
attractive color-Coulomb interaction with a long-range confining
interaction provides an adequate account of the interaction between a
quark and antiquark.  While the non-relativistic potential model was
originally introduced for heavy quarkonium systems, relativistic and
non-relativistic quark models using constituent quark masses have been
used to describe mesons with one or two light quarks as
quark-antiquark bound states \cite{God85,Bar92,Won02,Cra04,Cra06}.

Potential models have also been used to study heavy quarkonium bound
states at high temperatures
\cite{Mat86,Dig01a,Won01a,Kac02,Shu03,Won05,Won05a,Won06,Won06a,Son05,Alb05,Bla05,Man05,Dig05}.
At temperatures above the phase transition temperature, the potential
between a quark and an antiquark is subject to screening \cite{Mat86}.
Heavy quarkonium states become unbound in the potential as temperature
rises.  Potential models provide beneficial complements to lattice
gauge calculations as potential models allow simple and intuitive
evaluations of many quantities, some of which may be beyond the scope
of present-day lattice gauge calculations.  The central question in the
potential model has been focused on the characteristics and the
temperature dependencies of the screening potential as determined by
lattice gauge calculations
\cite{Mat86,Dig01a,Won01a,Kac02,Shu03,Won05,Won05a,Won06,Won06a,Son05,Alb05,Bla05,Man05,Dig05}.

There is recently a serious theoretical question whether it is
appropriate to apply a potential model to study heavy quarkonia at
high temperatures \cite{Moc06,Moc06a}.  On the one hand, lattice gauge
spectral function analyses have been carried out to investigate the
stability of heavy quarkonia at high temperatures
\cite{Asa03,Mat01,Dat03,Pet05,Jak06}.  On the other hand, independent
lattice gauge calculations \cite{Kac03,Kac05} have been performed to
evaluate various thermodynamical quantities, such as the free energy
$F_1({\bf r},T)$ and the internal energy $U_1({\bf r},T)$, for a
color-singlet static $Q$-$\bar Q$ pair at various separations ${\bf
r}$ and temperatures $T$. It remains an important open theoretical
question how a quark-antiquark potential can be extracted from these
thermodynamical quantities. Various potential models have been
proposed, utilizing $F_1$ \cite{Dig01a,Won01a,Bla05}, $U_1$
\cite{Kac02,Shu03,Alb05,Man05}, or a linear combination of $F_1$ and
$U_1$ with coefficients that depend on the equation of state
\cite{Won05,Won05a,Won06,Won06a}.  Although the latter
linear-combination model has been justified by theoretical arguments
and leads to dissociation temperatures that are consistent with
lattice gauge spectral function dissociation temperatures
\cite{Won05,Won05a,Won06,Won06a}, it remains necessary to confront the
model with other lattice gauge data to assess the degree of its
usefulness.

The spectral function is related to the meson current-current
correlator by a generalized Laplace transform.  In principle, they
carry equivalent information on the composite system.  One would
expect intuitively that the consistency of the lattice gauge spectral
function dissociation temperatures with the potential model
dissociation temperatures using the potential of \cite{Won05} would
lead to consistency of the lattice gauge correlators with the
corresponding potential model correlators.  Recent works in Refs.\
\cite{Moc06,Moc06a} however make the contrary claim that the meson
correlators obtained with many different types of potential models are
not consistent with lattice gauge correlators, and consequently
potential models cannot describe heavy quarkonia above $T_c$.

The failure of the potential model correlators to reproduce lattice
gauge correlators in Refs.\ \cite{Moc06,Moc06a} {\it for all cases}
suggests that the lack of agreement may not be due to the potential
models themselves but to the method of evaluating the meson
correlators in the potential model.  In the work of Ref.\
\cite{Moc06,Moc06a}, continuum states arising from a free fermion $Q$
and $\bar Q$ pair in a fermion gas contribute to the correlator.
However, to be able to propagate as a composite meson so as to
contribute to the correlator, the quark and the antiquark must be
temporally and spatially correlated to be a composite object with a
sufficiently long lifetime.  Continuum states in the free quark and
antiquark gas may not have sufficient temporal and spatial
correlations to be composite objects for such a propagation.

While we raise questions on the method of evaluating the potential
model correlators in \cite{Moc06,Moc06a}, we wish to present what we
view as a proper treatment of the current-current correlator in the
potential model.  We wish to point out the basic principle that out of
the continuum states, only resonance states and Gamow states
\cite{footnote} with lifetimes of sufficient magnitude can propagate
as composite objects and contribute to the meson current-current
correlator.  With the simple example of the pseudoscalar correlator,
we shall show in this paper that when both the bound state and the
continuum state contributions are properly treated, the potential
model of \cite{Won05} using a linear combination of $F_1({\bf r},T)$
and $U_1({\bf r},T)$ yields correlators consistent with lattice gauge
correlators, while the $F_1$ potential and the $U_1$ potential lead to
deviations.  Our results indicate consistency of the potential model
of \cite{Won05} with both the lattice gauge spectral function analyses
and the lattice gauge correlator analyses. The potential model of
\cite{Won05} thus gains support to be a useful tool to complement
lattice gauge calculations for the study of $Q\bar Q$ states at high
temperatures.

In Section II, we review the basic assumptions in the treatment of the
continuum states in Refs.\ \cite{Moc06,Moc06a}. We review in Section
III the  relationship between the current-current correlator and the
quarkonium wave functions.  In Section IV, we show how the resonance
states and Gamow states are characterized in the potential model. In
Section V, the potential model correlator is expressed as a sum over
contributions from meson bound and continuum states with lifetimes
greater than a certain limit.  In Section VI, we discuss the relation
between the $Q$-$\bar Q$ potential and lattice gauge thermodynamic
quantities.  In Section VII, we evaluate the potential model
correlators and find that the correlators obtained with the potential
of \cite{Won05} have features similar to those of lattice gauge
correlators.  We have thus demonstrated the consistency of the
potential model correlators with lattice gauge correlators.  In
Section VIII, we discuss the implications of the potential model
analysis on the assume default spectrum in the lattice gauge spectral
function analysis.  In Section IX we present our conclusions.

\section{Meson Correlators and Continuum States}

The meson (current-current) correlators $G(\tau,{\bf X})$ is a
function of the Euclidean time $\tau$ and the spatial coordinate ${\bf
X}$ defined by
\begin{equation}
\label{eq1}
G(\tau ,\mathbf{X})=\langle J_{M}(\tau ,\mathbf{X})J_{M}^{\dagger }(0,
\mathbf{0})\rangle,
\end{equation}
where $J_M (\tau,{\bf X})=\bar q (\tau,{\bf X})\Gamma_M q(\tau,{\bf
  X})$ and $\Gamma_M=1,\gamma^5,\gamma^\mu,\gamma^5\gamma^{\mu}$ are
the operators appropriate for scalar, pseudoscalar, vector, and
axial-vector mesons respectively. It is the probability amplitude for
creating a meson $M$ at space-time point $(0,{\bf 0})$, propagating
the meson to $(\tau,{\bf X})$, and destroying the meson at $(\tau,{\bf
  X})$.  Specializing to the case of the meson momentum ${\bf P}$
equal to ${\bf 0}$, the meson correlator depends then on the meson mass
spectrum $\sigma(\omega,T)$.

In their test of the potential model, the authors of
\cite{Moc06,Moc06a} assume that the meson mass spectrum in the
potential model is given by
\begin{eqnarray}
\label{sig}
\sigma(\omega,T)=\sum_i 2M_i F_i \delta(\omega^2-M_i^2)+\frac{4}{8\pi^2}
\theta(\omega-s_0)(a_H+b_H \frac{s_0^2}{\omega^2})
\sqrt{1-\frac{s_0^2}{\omega^2}},
\end{eqnarray}
where $M_i(T)$ is a bound state meson mass calculated in the potential
model, $F_i(T)$ is the corresponding magnitude of the wave function at
the origin.  The second term with the step function represents
continuum meson states obtained by assuming that the $Q$ and $\bar Q$
are free fermions in a fermion gas \cite{Kar01}.  The quantity
$s_0(T)$ is the continuum threshold, $a_H$ and $b_H$ are constants
that depend on the meson type.  The potential model meson correlator
$G(\tau)$ is then obtained by folding the meson mass spectrum
$\sigma(\omega,T)$ with the propagating kernel $K(\tau,\omega,T)$,
\begin{eqnarray}
G(\tau)=\int d\omega \sigma(\omega,T) K(\tau,\omega,T),
\end{eqnarray}
where $K(\tau,\omega,T)$ is given by
\begin{eqnarray}
K(\tau,\omega, T)=\frac{\cosh [\omega(\tau -1/2T)]}{\sinh [ \omega/2T]}.
\end{eqnarray}
The lattice gauge correlator $G(\tau)$ is represented relative to the
``reconstructed'' correlator $G_{\rm recon}(\tau)$ calculated with the
meson mass spectrum at $T=0$ \cite{Dat03,footnote2}.  The lattice
gauge pseudoscalar correlators obtained in \cite{Dat03} are shown in
Fig.\ 1a.  The potential model correlators $G(\tau)/G_{\rm
recon}(\tau)$, obtained in \cite{Moc06,Moc06a} with a screening
potential (Fig. 1b) or with the potential of Ref.\ \cite{Won05,Won06}
(Fig. 1c), are found to be qualitatively very different from the
lattice gauge correlators of Fig. 1a.  Significant differences occur
for both charmonia and bottomia, at all temperatures above $T_c$, and
for many fundamentally different potentials.  The authors of Ref.\
\cite{Moc06,Moc06a} then draw the conclusion that the potential model
is not a good description for heavy quarkonia above $T_c$.

\vspace*{-2.7cm}
\begin{figure} [h]
\includegraphics[angle=0,scale=0.80]{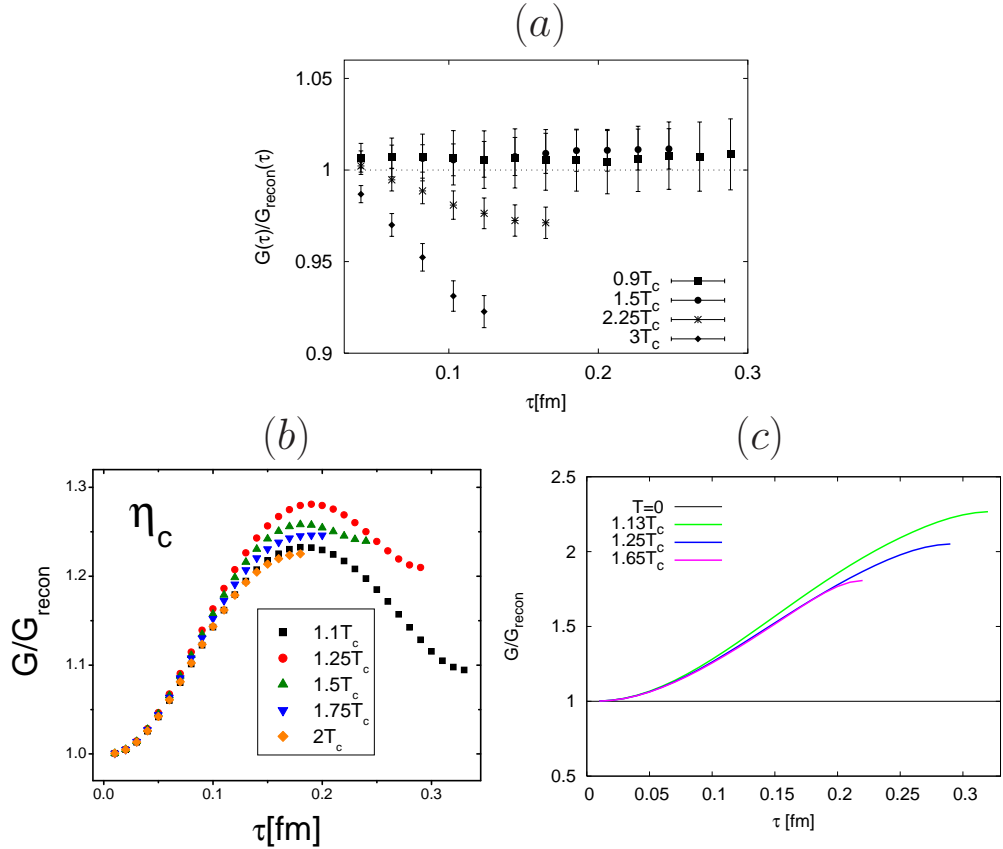}
\vspace*{-9.0cm}
\caption{ Fig. 1a shows the pseudoscalar charmonium correlators
  $G(\tau)/G_{\rm recon}(\tau)$ obtained in lattice gauge
  calculations \cite{Dat03,footnote2}.
  Fig. 1b shows the potential
  model correlators for the pseudoscalar charmonium obtained in Ref.\
  \cite{Moc06} with the screened Cornell potential.  Fig. 1c gives the
  potential model correlators for the pseudoscalar charmonium obtained in
  Ref.\ \cite{Moc06a} with the potential of Refs.\
  \cite{Won05,Won06}.}
\end{figure}

In the evaluations of the potential model correlators in
\cite{Moc06,Moc06a}, contributions from continuum states represented
by the second term in Eq.\ (\ref{sig}) have been included.  While the
spectrum in the continuum states is continuous as in Eq.\ (\ref{sig}),
not all continuum states represented by the step function $Q\bar Q$
free-fermion gas states in the second term in Eq.\ (\ref{sig}) possess
the proper characteristics to propagate as a composite meson so as to
contribute to the meson correlator $G(\tau)$.  To be able to
propagate as a composite meson from time 0 to $\tau$ so as to
contribute to the correlator $G(\tau)$, the quark and antiquark must
be temporally and spatially correlated.  The composite object must
have a composite lifetime sufficiently long compared to the
propagation time $\tau$.  Continuum states in the free quark and
antiquark gas may not have sufficient temporal and spatial
correlations between $Q$ and $\bar Q$ to qualify as being composite
objects for such a propagation.

A similar question poses itself in low-energy nuclear physics in the
evaluation of the density of continuum states of a composite object
formed by a particle (neutron, say) and a nucleus (represented by a
potential well).  States for the free particle in a (neutron) fermion
gas do not have relative temporal and spatial correlations between
the particle and the nucleus to qualify as being a composite object
formed by the particle and the nucleus in the continuum.  It is for
this well-known reason that to get the level density of continuum
states of a composite object formed by a particle and a nucleus, it is
necessary to subtract the level density of free gas states from the
total density of continuum states
\cite{Tub79,Shl92,Shl95,Shl97,Cha05,Alh03}.  In the analogous problem
of level density of continuum states of a composite $Q\bar Q$ meson,
one can carry out the free fermion gas states subtraction or
alternatively use only resonance states and Gamow states as
contributors to the continuum level density
\cite{Tub79,Shl92,Shl95,Shl97,Cha05,Alh03}.  Therefore, for our case
of meson correlators for a composite $Q\bar Q$ meson, the basic
principle is that the proper continuum meson states that can
contribute to the meson correlator $G(\tau)$ should be limited to
meson resonance states and Gamow states which have composite object
lifetime long compared to $\tau$.  By focusing our attention on
the pseudoscalar charmonium as an example, we would like to demonstrate
that the potential model with the correct set of bound and continuum
Gamow states can describe lattice gauge correlators.

\section{The Meson Correlator and Meson Internal Wave functions}

We start with the current-current correlator $G(\tau
,\mathbf{X})=\langle J_{M}(\tau ,\mathbf{X})J_{M}^{\dagger }(0,
\mathbf{0})\rangle$ of Eq.\ (\ref{eq1}) and restrict ourselves to the
consideration of the pseudoscalar charmonium.  The current
operator is just $J_{M}(\tau ,\mathbf{X})=\bar{q}(\tau ,\mathbf{X}
)\gamma _{5}q(\tau ,\mathbf{X})$.  The field operator has both
annihilation and creation parts, $q(\tau ,\mathbf{X})=q^{(+)}(\tau ,
\mathbf{X})+q^{(-)}(\tau ,\mathbf{X})$. \ Hence, the current operator
can be decomposed into
\begin{equation}
J_{M}(\tau ,\mathbf{X})=\bar{q}^{(-)}(\tau ,\mathbf{X})\gamma
_{5}q^{(+)}(\tau ,\mathbf{X})+\bar{q}^{(+)}(\tau ,\mathbf{X})\gamma
_{5}q^{(+)}(\tau ,\mathbf{X})+\bar{q}^{(-)}(\tau ,\mathbf{X})\gamma
_{5}q^{(-)}(\tau ,\mathbf{X})+\bar{q}^{(+)}(\tau ,\mathbf{X})\gamma
_{5}q^{(-)}(\tau ,\mathbf{X}).  \label{q}
\end{equation}

We consider the propagation of the pseudoscalar meson $M$ from the
initial (Euclidean) space-time coordinate $(0,\mathbf{0})$ to the
final space-time coordinate $(\tau , \mathbf{X})$. It is often
convenient to study
\begin{equation}  \label{cor}
G(\tau )=\int d\mathbf{X}G(\tau ,\mathbf{X})=\int d\mathbf{X}\langle
J_{M}(\tau ,\mathbf{X})J_{M}^{\dagger }(0,\mathbf{0})\rangle .
\end{equation}
The temporal behavior of the correlator $G(\tau )$ provides information on
the meson spectrum.

For the extraction of information on the correlator we focus only on
the particle-antiparticle creation operator portion
$\bar{q}^{(+)}(\tau , \mathbf{X})\gamma _{5}q^{(-)}(\tau
,\mathbf{X})~$ of Eq.\ (\ref{q}).  The $ \gamma _{5}$ matrix singles
out the dominant lower component of the operator $ q^{(-)}(\tau
,\mathbf{X})~$\ for the antiquarks.  We rename that dominant component
by using the notation
\begin{equation}
J_{M}(\tau ,\mathbf{X})=\psi _{1}(x_{1})\psi _{2}(x_{2}){\biggr |}
_{x_{1}=x_{2}=(\tau ,\mathbf{X})},
\end{equation}
where particle 1 represents $Q$ and particle 2 represents $\bar{Q}$,
$\psi _{1}(x_1)$ and $\psi_2(x_2)$ are the dominant $Q$ and $\bar Q$
fields.  The meson current-current correlator of Eq. (\ref{eq1})
becomes
\begin{equation}
\label{rr}
G(\tau ,\mathbf{X})=\langle ~\psi_{1}(x_{1})\psi _{2}(x_{2})
 \psi _{1}^\dagger (x_{1}') \psi
_{2}^\dagger (x_{2}') ~\rangle{\biggr |} _{x_{1}=x_{2}=(\tau
;\mathbf{X}); x_{1}'=x_{2}'=(0 ,\mathbf{0})}.
\end{equation} 
The above current $J_M(\tau ,\mathbf{X})$ is a local current where
$(\tau ,\mathbf{X})=X$ is the CM coordinate $X$ of the $Q$-$\bar{Q}$
pair with $X=(x_{1}+x_{2})/2$, and the relative coordinate,
$x=(x_{1}-x_{2})$, is restricted to be zero.  A general current
containing a more general relative coordinate $x=(x_{1}-x_{2})$ is
\begin{equation}
J_{M}(X,x)=\psi _{1}(X+x/2)\psi _{2}(X-x/2),
\end{equation} 
with the more general Green's function
\begin{equation}
G(x_{1},x_{2};x_{1}^{\prime },x_{2}^{\prime })
=\langle \psi_{1}(x_{1})\psi _{2}(x_{2})
 \psi _{1}^\dagger (x_{1}') \psi
_{2}^\dagger (x_{2}') \rangle.
\end{equation} 
We would like to make a transformation of the coordinate system from
the separate particle coordinates $(x_{1},x_{2})$ to $(X,x)$ and
introduce the field operator $\hat{\Psi}_{M}(X)$ to represent the field of
a composite meson which has the internal relative motion between the quark
and the antiquark described by an internal wave function $\psi _{M}$.

Following \cite{Won00}, we define the field operator
${\hat{\Psi}}_{M}(X)$ for CM motion by
\begin{equation}
{\hat{\Psi}}_{M}(X)|X\rangle =\Psi _{M}(X)|X\rangle ,
\end{equation} 
with the wave function $\Psi _{M}(X)$ satisfying Eq. (2.23) of \cite{Won00} 
\begin{equation}
\left\{ P^{2}-M^{2}\right\} \Psi _{M}(X)=0, 
\end{equation} 
where $P=p_1+p_2$ is the total momentum of $Q$ and $\bar Q$, $p_1$ is
the momentum of of $Q$, and $p_2$ is the momentum of $\bar Q$.
For each composite state of mass $M$, there is a wave function
${{\psi}}_{M}(x_\perp)$ for relative motion satisfying Eq.
(2.24) of \cite{Won00}
\begin{equation}
\left\{ p^{2}-\Phi (x_{\perp },M)+b^{2}(M^{2},m_{1}^{2},m_{2}^{2})\right\}
\psi _{M}(x_{\perp })=0,  \label{eq2}
\end{equation} 
where $p=(p_{1}-p_{2})/2$ is the relative momentum of $Q$ and $\bar
Q$,  $\Phi (x_{\perp },M)$ is the two-body interaction
potential that acts between $Q$ and $\bar{Q}$, and
\begin{eqnarray}
x_{\perp } &=&x-\frac{PP\cdot x}{P^{2}}, \\
P\cdot x_{\perp } &=&0.
\end{eqnarray} 
The composite particle mass $M$ is the constant of the separation of
variables and is related to the eigenvalue $b^2$ of the
Schr\" odinger-type equation for relative motion, Eq.\ (\ref{eq2}), by
\begin{equation}
b^{2}(M^{2},m_{1}^{2},m_{2}^{2})=\frac{1}{4M^{2}} 
\{ M^{4}-2M^{2}(m_{1}^{2}+m_{2}^{2})+(m_{1}^{2}-m_{2}^{2})^{2}\}.
\end{equation} 

We decompose the current field operator as a sum
of composite meson operators ${\hat \Psi} _{n}(X)$ with coefficients
$\psi _{n}(x_{\perp })$,
\begin{equation}
J_{M}(X,x_{\perp })=\sum_{n}\psi _{n}(x_{\perp }){\hat  \Psi} _{n}(X),
\end{equation} 
in which the summation $n$ includes a sum over bound and continuum
states. \ As the field operators of composite objects of different
types and energies produce orthogonal states, we get
\begin{eqnarray}
G(x_{1},x_{2};x_{1}^{\prime },x_{2}^{\prime })
=G(X,X^{\prime }; x_{\perp },x_{\perp }^{\prime })
=\sum_{n} \psi _{n}(x_{\perp }) \psi
_{n}^*(x_{\perp }^{\prime }) \langle {\hat \Psi} _{n}(X)
{\hat \Psi} _{n}^{\dagger}(X^{\prime
})\rangle .
\end{eqnarray} 
The Green's function for a composite meson of mass
$M_n$ is
\begin{eqnarray}
g_n(X,X^{\prime })
=\langle {\hat \Psi} _{n}(X){\hat \Psi}_{n}^\dagger (X^{\prime
})\rangle ,
\end{eqnarray} 
we have therefore
\begin{eqnarray}
\label{bigG}
G(X,X^{\prime };x_{\perp}, x_{\perp }^{\prime })=\sum_{n}
\psi_{n}(x_{\perp })\psi_n^*(x_{\perp }^{\prime })
g_{n}(X,X^{\prime }).
\end{eqnarray} 
The advantage of the separation of the correlation function in terms
of the relative degrees of freedom $\{x_{\perp}, x_{\perp }^{\prime
}\}$ and the composite particle CM degrees of freedom
$\{X,X^{\prime }\}$ is that the Green's function $g_n(X,X')$ corresponds
to that of a free single particle (meson) of mass $M_{n}$ in a thermal
bath of the plasma \cite{Kad62}. We consider $X^{\prime }=0$ and
$X=(\tau ,\mathbf{X)}$ with $ 0\leq \tau \leq \beta=1/T $ and take the
spatial Fourier transform of $g_{n}(\tau {\bf X},0)$ with respect to
${\bf X}$,
\begin{eqnarray}
{\tilde g}_{n}(\tau, \mathbf{P})=\int d\mathbf{X}e^{-i\mathbf{P}\cdot
\mathbf{X} }g_{n}(\tau \mathbf{X},0).
\end{eqnarray}
The correlator of a free particle with $0\leq \tau \leq \beta $ is given by
[see e.g. the equation before (3.4) of Ref.\ \cite{Kad62}] 
\begin{eqnarray}
\label{exp}
{\tilde g}_{n}(\tau,\mathbf{P} )=\int \frac{dP_{0n}}{2\pi i}e^{-P_{0n}\tau }\frac{A( 
\mathbf{P},P_{0n})}{1-\exp \{-\beta (P_{0n}-\mu )\}},
\end{eqnarray} 
where $\mu $ is the chemical potential for the composite particle in
the medium and $A(\mathbf{P},P_{0})$ is the spectral function for a
free composite particle given by [see e.g. equation after (3-8b) of
\cite{Kad62} for the non-relativistic case],
\begin{eqnarray}
A({\bf P},P_{0n})=2\pi \delta (P_{0n}-\sqrt{\mathbf{P}^{2}+M_{n}^{2}}).
\end{eqnarray}
The Green's function for the composite meson in the state $n$ with ${\bf
P}=0$ is then
\begin{eqnarray}
{\tilde g}_n(\tau)\equiv {\tilde g}_{n}(\tau, \mathbf{P}=0 )=
\int d\mathbf{X} ~ {\tilde g}_{n}(\tau \mathbf{X})
=\frac {1}{i}
 \frac{\exp\{-M_{n}\tau \}} {1-\exp \{-\beta (M_{n}-\mu )\}}
\end{eqnarray} 
The corresponding correlator $G(\tau)$ in Eq.\ (\ref{cor}) and (\ref{bigG}) 
is then
\begin{eqnarray}
G(\tau; x_{\perp},x_{\perp}^{\prime })
=\frac{1}{i}\sum_{n}
\psi _{n}(x_\perp,M_{n}) \psi _{n}^*(x_\perp^{\prime },M_{n}) 
\frac{\exp (-M_{n}\tau )}{1-\exp \{-\beta (M_{n}-\mu )\}}.
\end{eqnarray}
In the center-of-mass system of the composite particle at rest,
$x_\perp=(0,{\bf r})$, and we have
\begin{eqnarray}
G(\tau; {\bf r},{\bf r}^{\prime })
=\frac{1}{i}\sum_{n}
\psi _{n}({\bf r},M_{n}) \psi _{n}^*({\bf r}^{\prime },M_{n}) 
\frac{\exp (-M_{n}\tau )}{1-\exp \{-\beta (M_{n}-\mu )\}}.
\end{eqnarray}
In the above equation, the propagator kernel for a composite particle
in state $n$ with a mass $M_n$ can be re-written as
\begin{eqnarray}
{\tilde g}_n(\tau) = \frac{\exp (-M_{n}\tau )} {1-\exp \{-\beta (M_{n}-\mu )\}}
=\frac{\exp \{ -M_n(\tau-\beta/2)+\beta\mu/2 \} } 
{2 \sinh \{\beta(M_n-\mu)/2\}}.
\end{eqnarray}

In lattice gauge calculations, one chooses to work with the case of
$\mu=0$ and imposes the periodic boundary condition,
$G(\tau)|_{\tau=0} = G(\tau)|_{\tau=\beta}$.  The periodic boundary
condition can be satisfied by including not only the exponentially
decreasing solution $e^{-M_{n}\tau}$ but also the exponentially
increasing solution $e^{M_{n}\tau}$ in Eq.\ (\ref{exp}) so that the
propagating kernel becomes,
\begin{eqnarray}
\label{kw}
K(\tau,M_n,T) 
=\frac{\exp \{ -M_n(\tau-\beta/2) \}
      +\exp \{  M_n(\tau-\beta/2) \}
 } {2 \sinh \{\beta M_n/2\}}
=\frac{\cosh\{ M_n(\tau-\beta/2)\}}{\sinh \{\beta M_n/2\}},
\end{eqnarray}
which satisfies the periodic boundary condition,
$K(0,M_n,T)=K(\beta,M_n,T)$.  We shall consider this case of $\mu=0$
with the periodic boundary condition and use the above propagating
kernel, in order to compare the potential model correlators with the
lattice gauge correlators.  We shall however compare results only for
$\tau$ less than and up to $\beta/2$ before the onset of the dominance
of the exponentially increasing evolution.  For the case considered in
lattice gauge calculations, where the relative coordinate of the
$Q$-$\bar{Q}$ pair is set to zero, we have
$\mathbf{r}=\mathbf{r}^{\prime }=\mathbf{0}$ and thus
\begin{eqnarray}
\label{GGG}
G(\tau) \equiv G(\tau; {\bf r}, {\bf r}^\prime)
\biggr |_{{\bf r}={\bf r}'={\bf 0}}
=\frac{1}{i} \sum_{n}K(\tau,M_n, T) 
\psi_{n}(\mathbf{r,}M_{n})\psi _{n}^*(\mathbf{r}^{\prime },M_{n}) 
\biggr |_{{\bf r}={\bf r}'={\bf 0}} .
\end{eqnarray}

\section{Gamow States and Resonance States in the Continuum}

We would like to use Eq.\ (\ref{GGG}) to evaluate correlators for
the pseudoscalar charmonium in the potential model, for comparison
with pseudoscalar correlators obtained in lattice gauge
calculations.  As the mass of a charm quark is quite large, we shall
restrict ourselves to non-relativistic considerations.  For
simplicity, we shall also ignore spin.  The summation over $n$ can be
separated into a sum over bound states $b$ and a sum over continuum
states ${\bf k}$.  The proper treatment of the contributions from the
continuum states is important in the evaluation of the correlator.

As we remarked earlier, the meson correlator $G(\tau ,\mathbf{X})$ is
the probability amplitude for creating a meson $M$ at space-time point
$(0,{\bf 0})$, propagating the meson to $(\tau,{\bf X})$, and
destroying the meson at $(\tau,{\bf X})$.  To be able to propagate as
a composite object from time 0 to $\tau$, the quark and antiquark pair
must be spatially and temporally correlated.  The composite object
must have a lifetime long compared to the propagation time $\tau$.
Using considerations similar to those in the density of states of a
composite object in nuclear physics
\cite{Tub79,Shl92,Shl95,Shl97,Cha05,Alh03}, we are well advised that
the proper continuum states that can contribute to the meson
correlator $G(\tau,{\bf X})$ should be resonance states and Gamow
states with lifetimes long compared to $\tau$.

We need information on the lifetime of a continuum state ${\bf k}$ in
the potential model.  For that purpose, we can calculate the phase
shift $\delta(\epsilon_{\bf k})$ as a function of the continuum state
energy $\epsilon_{\bf k}={\bf k}^2/2\mu_m$ where $\mu_m$ is the
reduced mass, $m_Q/2$.  Knowing the phase shift as a function of
energy, we can determine the delay time, as pointed out by Wigner
\cite{Wig55, New66,Des74,Sat90}.  To obtain such a relationship, we
consider the wave packet with momenta centered around ${\bf k}$ with
energy $\epsilon_{\bf k}$ to travel from ${\bf r}=0$ to ${\bf R}$ with
${\bf R}$ along ${\bf k}$, as a function of $t$.  The peak of the wave
packet arises from the interference of two waves with energy
$\epsilon_{\bf k}$ and $\epsilon_{{\bf k}'}$.  They must interfere
constructively.  Constructive interference is possible when the phase
difference of the two wave functions at $\epsilon_{\bf k}$ and
$\epsilon_{{\bf k}'}$ is zero. This condition for the constructive
interference can be written explicitly as
\begin{eqnarray}
{\bf k}\cdot {\bf R}-{\bf k}'\cdot {\bf R}+
\delta(\epsilon_{\bf k})-\delta(\epsilon_{{\bf k}'})
-(\epsilon_{\bf k} t-\epsilon_{{\bf k}'} t)=0.
\end{eqnarray}
Solving for $t$ and taking the limit ${{\bf k}'}$ approaches ${{\bf
k}}$, we obtain
\begin{eqnarray}
t=\frac{R}{d\epsilon_{\bf k}/d{ k}}
+\frac{\partial \delta(\epsilon_{\bf k})}{\partial \epsilon_{\bf k}},
\end{eqnarray}
which indicates that the passage of the wave packet with momentum
centered at ${\bf k}$ from the origin to ${\bf R}$ is delayed by a delay time
\cite{Wig55}
\begin{eqnarray}
  ({\rm delay~time}) = \frac{\partial \delta(\epsilon_{\bf k})}
{\partial \epsilon_{\bf k}}.
\end{eqnarray}
A negative delay time, with a negative $\partial \delta(\epsilon_{\bf
  k})/{\partial \epsilon_{\bf k}}$, represents the flying apart of the
$Q$ and $\bar Q$ in advance of their coalescence approach and by
causality cannot represent a composite object.  Continuum states with
negative $\partial \delta(\epsilon_{\bf k})/{\partial \epsilon_{\bf
    k}}$ should not be included as contributing to the meson
correlator for the propagation of the composite quarkonium.

A system with a positive delay time can be interpreted as possessing a
finite lifetime.  What fraction of the delay time should be assigned
to the lifetime of this composite object residing as a wave packet
centered around the continuum state ${\bf k}$?  To answer such a
question, we seek the help of the case of a sharp resonance for which
the answer can be readily obtained.

If the state at ${\bf k}$ is a sharp resonance, the phase shift
$\delta_{\bf k}(\epsilon)$ in the neighborhood of the resonance ${\bf
   k}$ with a small width $\Gamma_{\bf k}$ is given by
\begin{eqnarray}
\tan \delta_{\bf k}(\epsilon) = \frac {\Gamma_{\bf k}/2}{\epsilon_{\bf
k}-\epsilon}.
\end{eqnarray} 
Then the derivative of the phase shift with respect to the continuum
energy is
\begin{eqnarray}
\frac{\partial \delta_{\bf k}(\epsilon)} {\partial \epsilon} \sim
\frac{\Gamma_{\bf k}/2}  {(\epsilon_{\bf k}-\epsilon)^2 
+[\Gamma_{\bf k}/2]^2}.
\end{eqnarray}
At the resonance $\epsilon=\epsilon_{\bf k}$, we have
\begin{eqnarray}
\frac{\partial \delta_{\bf k}(\epsilon)}{\partial \epsilon} 
\biggr | _{\epsilon=\epsilon_{\bf k}} 
\sim \left [ \frac{\Gamma_{\bf k}}{2}\right ] ^{-1}=({\rm delay~time}).
\end{eqnarray}

Thus, by examining the case of a sharp resonance state, we find that
half of the delay time can be assigned to the lifetime, $1/\Gamma_{\bf
k}$, of the continuum state. We shall therefore assume that the
assignment of the fraction of one-half of the delay time as the
composite particle lifetime is a reasonable concept, for all states
with positive $\partial \delta_{\bf k}(\epsilon_{\bf k})/\partial
\epsilon_{\bf k}$.  Thus, we assign the width $\Gamma_{\bf k}$
associated with a state ${\bf k}$ with a time delay as
\begin{eqnarray}
\label{gamm}
\Gamma_{\bf k}=\frac{2}{\partial \delta_{\bf k}(\epsilon_{\bf k})/\partial
\epsilon_{\bf k}}.
\end{eqnarray}

A well-defined potential resonance in the continuum with momentum
${\bf k}$ and energy $\epsilon_{\bf k}$ occurs when $\partial
\delta_{\bf k}(\epsilon_{\bf k})/\partial \epsilon_{\bf k} >0$ and
$\delta_{\bf k}(\epsilon_{\bf k})=(2n+1)\pi/2$, where $n$ is an
integer.  States at energies with $\delta_{\bf k}(\epsilon_{\bf
k})=(2n+1)\pi/2$ but with $\partial \delta_{\bf k}(\epsilon_{\bf
k})/\partial \epsilon_{\bf k} <0$ are echos and not resonances
\cite{Mcv67}.  For these reasons, if the $S$-wave potential does not
possess a barrier to trap the waves in the interior, there will be no
$S$-wave resonances \cite{New66}. Nevertheless at plasma temperatures
above $\sim$1.6$T_c$ (as we shall see in subsequent sections), there
are $S$-wave continuum states with positive time delays and lifetimes
and can be represented by Gamow states with various widths.  They will
be used in our subsequent calculations of the potential model
correlators.

\section{Meson Correlator in the Potential Model}

Upon limiting our attention in this manuscript to the pseudoscalar
charmonium with $L$=0, we note that the quark-antiquark potential
itself does not possess a potential barrier, and thus there are no
sharp $S$-wave resonances.  There are however continuum states with
positive time delays at $T>\sim 1.6 T_c$, as we shall see in Section VII.
These continuum states can be represented by Gamow states with finite
lifetimes and widths.

With composite particle states consisting of bound states and Gamow
states, the summation over $n$ in Eq.\ (\ref{GGG}) for the evaluation
of the correlator in the potential model can be separated into a sum
over bound states $b$ and an integral over continuum Gamow states ${\bf k}$,
\begin{eqnarray}
\label{sum0}
\sum_n  \psi_{n}({\bf r},M_{n})
        \psi _{n}^*({\bf r}^{\prime },M_{n})  f(M_n)
&\to&  
\int d\omega \sum_{b}\delta ({\omega -M_{b}})
        \psi_{b}({\bf r},M_b))
        \psi _{b}^*({\bf r}^{\prime },M_{b})  f(\omega)
\nonumber\\
&+& \int_{0}^{\infty }  d\epsilon \int  d{\bf k} 
D_{\bf k}(\epsilon,\Gamma_{\bf k})
\theta(\partial \delta_{\bf k}/ \partial \epsilon_{\bf k})
\psi_{{\bf k}}({\bf r}) \psi_{{\bf k}}^*({\bf r}^{\prime })
f(\epsilon) 
\end{eqnarray}
where the width of the Gamow state $\Gamma_{\bf k}$ is $2/ [\partial
\delta_{\bf k}(\epsilon_{\bf k})/ \partial \epsilon_{\bf k})]$ and the
function $D_{\bf k}(\epsilon,\Gamma_{\bf k})$ represents the spreading
of the distribution of the Gamow states ${\bf k}$ in continuum energy
$\epsilon$ due to the presence of its delay time and a width
$\Gamma_{\bf k}$.  One can choose different forms of the distribution
function $D_{\bf k}(\epsilon,\Gamma_{\bf k})$ that has a peak at
$\epsilon_{\bf k}$ with a width $\Gamma_{\bf k}$, such as a Gaussian
or a Breit-Wigner distribution in $\epsilon$.  The results in the
correlators would not be sensitive to the form of the distribution
function $D_{\bf k}(\epsilon,\Gamma_{\bf k})$.  For convenience, we
shall choose to represent $D_{\bf k}(\epsilon,\Gamma_{\bf k})$ by the
Breit-Wigner distribution
\begin{eqnarray}
D_{\bf k}(\epsilon,\Gamma_{\bf k})
=\frac{a_k}{\pi}
\frac{\Gamma_{\bf k}/2}
{(\epsilon-\epsilon_{\bf k})^2 +(\Gamma_{\bf k}/2)^2 },
\end{eqnarray}
where
\begin{eqnarray}
a_k=\frac{2\pi}{\pi+2 \arctan (2\epsilon_{\bf k}/\Gamma_{\bf k})},
\end{eqnarray}
and $\int_0^{\infty} d\epsilon D_{\bf k}(\epsilon,\Gamma_{\bf k})=1$.
Eq.\ (\ref{sum0}) becomes
\begin{eqnarray}
\label{sum}
\sum_n  \psi_{n}({\bf r},M_{n})
        \psi _{n}^*({\bf r}^{\prime },M_{n})  f(M_n)
&\to&  
\int d\omega \sum_{b}\delta ({\omega -M_{b}})
        \psi_{b}({\bf r},M_b))
        \psi _{b}^*({\bf r}^{\prime },M_{b})  f(\omega)
\nonumber\\
&+& \int_{0}^{\infty }  d\epsilon \int  d{\bf k} 
\frac{a_k}{\pi} 
\frac{(\Gamma_{\bf k}/2)
\theta(\partial \delta_{\bf k}/ \partial \epsilon_{\bf k})}
{(\epsilon-\epsilon_{\bf k})^2 +(\Gamma_{\bf k}/2)^2 }
\psi_{{\bf k}}({\bf r}) \psi_{{\bf k}}^*({\bf r}^{\prime })
f(\epsilon) 
\end{eqnarray}
We normalize the wave function of the bound state $M_b$ according to
\begin{eqnarray}
\int d{\bf r} |\psi_{b}(\mathbf{r},M_b ) |^2 =1.
\end{eqnarray}
The wave function in the continuum $\psi_{{\bf k}}({\bf r} )$
is normalized according to
\begin{eqnarray}
\int d{\bf r} ~~\psi_{{\bf k}}({\bf r})~ 
 \psi_{{\bf k}^{\prime}}^*({\bf r})=\delta({\bf k}-{\bf k}'), 
\end{eqnarray}
and it behaves asymptotically as 
\begin{eqnarray}
\label{asym}
\psi _{{\bf k}}({\bf r}\to \infty ) \to 
\frac{ \exp\{ i{\bf k}\cdot {\bf r}  \}}{(2\pi)^{3/2}}.
\end{eqnarray}

It is not sufficient to limit the width $\Gamma_{\bf k}$ to be
positive.  To be able to propagate temporally as a composite particle
from $0$ to $\tau$, the composite object must have a lifetime
exceeding a certain limit $\hbar/ \Gamma_{\rm max}$.  It is therefore
necessary to limit the contributions in the integral in Eq.\
(\ref{sum}) further by an additional step function $\theta(\Gamma_{\rm
max}-\Gamma_{\bf k})$.  The correlator is therefore
\begin{eqnarray}
G(\tau;{\bf r} {\bf r}' ) &=&
\frac{1}{i}\sum_{b}\frac
{\cosh \{M_b (\tau-\beta/2)\} } {\{\sinh \{M_\epsilon \beta/2\}}
\psi _{b}(\mathbf{r}, M_{b}) 
\psi _{b}^{\ast }(\mathbf{r}^{\prime },M_{b})
\nonumber\\ 
&+&\frac{1}{i} \int d\epsilon \int d{\bf k} \frac{a_k}{\pi}
\frac{(\Gamma_{\bf k}/2) 
\theta(\partial \delta_{\bf k}/\partial\epsilon_{\bf k})
\theta(\Gamma_{\rm max}-\Gamma_{\bf k})}
{(\epsilon-\epsilon_{\bf k})^2 +[\Gamma_{\bf k}/2]^{2}} 
\frac{\cosh \{ M_\epsilon (\tau-\beta/2)\} }
{\sinh \{M_\epsilon \beta/2\}}
\psi _{{\bf k}}(\mathbf{r}) \psi _{{\bf k}}^{\ast }(\mathbf{r}^{\prime }).
\end{eqnarray}

We need to specify the maximum width limit $\Gamma_{\rm max}$.  The
width limit $\Gamma_{\rm max}$ depends on the time scale in the
propagation of the composite meson. In lattice gauge calculations, the
composite object propagation time varies from $\tau=0$ to
$\tau=\beta/2$ with a maximum of $\tau=\beta/2$.  A propagation time
$\tau$ greater than $\beta/2$ will lead to the unphysical region where
the meson probability amplitude grows predominantly exponentially with
time (see Eq.\ (\ref{kw})).  With this maximum $\tau=\beta/2$ in the
correlator measurement, the composite object lifetime $\tau_{\rm
life}$ needs to be greater than $\beta/2$ and the minimum life time
$\tau_{\rm min~life}$ is given by
\begin{eqnarray}
\label{minlife}
\tau_{\rm min~life}(T)=\beta/2=1/2T.
\end{eqnarray}
This arises because a composite object with a lifetime $\tau_{\rm
life}$ less than $\tau_{\rm min~life}(T)$ will not be able to remain a
composite object as it propagates from $\tau=0$ to $\tau=\beta/2$.
For the continuum state at ${\bf k}$, the above requirement, that the
composite object lifetime $\tau_{\rm life} =1/\Gamma_{\bf k}$ must be
greater than or equal to $\tau_{\rm min~life}(T)=1/2T$, leads to the
condition for the maximum width $\Gamma_{\rm max}=1/\tau_{\rm
min~life}$ of the continuum states detectable by the correlator
measurement,
\begin{eqnarray}
\label{gammax}
\Gamma_{\rm  max}=\frac{1}{\tau_{\rm min~life}(T)}=\frac{1}{\beta/2}=2T.
\end{eqnarray}

For the case considered for the lattice gauge calculations where the
relative coordinate $x=(x_{1}-x_{2})$ of the $Q$-$\bar{Q}$ pair is set
to zero (Eq.\ (\ref{rr})), we have $\mathbf{r}=\mathbf{r}^{\prime
}=\mathbf{0}$ and thus we need to evaluate $G(\tau;{\bf r}{\bf r}')$
at ${\bf r}={\bf r}'={\bf 0}$.  We define the quantity ${\cal K}$ as
the the ratio of the absolute square of the amplitude of the continuum
wave function at the origin to the absolute square of its amplitude at
infinity \cite{Won97}
\begin{eqnarray}
{\cal K}({\bf k})=\frac{|\psi _{{\bf k}}(\mathbf{0} )|^2}
{|\psi _{{\bf k}}({\bf r}\to \infty )|^2},
\end{eqnarray}
which can be evaluated numerically for the potential in question.
Using Eq.\ (\ref{asym}), the continuum wave function at the origin is
therefore approximately
\begin{eqnarray}
|\psi _{{\bf k}}(\mathbf{0} )|^2=
{{\cal K}({\bf k})}/{(2\pi)^3}.
\end{eqnarray}
The correlator becomes
\begin{eqnarray}
\label{Gtau}
G(\tau )&\equiv& G(\tau;{\bf r} {\bf r}' )\biggr |_{{\bf r}={\bf r}'={\bf 0}}
=\frac{1}{i}\sum_{b}
\frac{\cosh \{ M_b (\tau-\beta/2)\} }
{\sinh \{M_b \beta/2\}}
\psi _{b}        (\mathbf{0}          ,M_{b})
\psi _{b}^{\ast }(\mathbf{0}^{\prime },M_{b})
\nonumber\\
&+& 
\frac{1}{i} \int d\epsilon \int d{\bf k} \frac{a_k}{\pi} 
\frac{(\Gamma_{\bf k}/2) 
\theta(\partial \delta_{\bf k}/\partial\epsilon_{\bf k})
\theta(\Gamma_{\rm max}-\Gamma_{\bf k}) }
{(\epsilon-\epsilon_{\bf k})^2 +[\Gamma_{\bf k}/2]^{2}} 
\frac{\cosh \{ M_\epsilon (\tau-\beta/2)\} }
{\sinh \{M_\epsilon \beta/2\}}
\frac{{\cal K}({\bf k})}{(2\pi)^3} .
\end{eqnarray}

\section{Relation between the $Q$-$\bar Q$ Potential and lattice Gauge
Thermodynamic Quantities }

In the potential model, the most important physical quantity is the
$Q$-$\bar Q$ potential between the quark $Q$ and the antiquark $\bar
Q$ in a color-singlet state. Previous works in the potential model use
the color-singlet free energy $F_1$ \cite{Dig01a,Won01a,Bla05} or the
color-singlet internal energy $U_1$ \cite{Kac02,Shu03,Alb05,Man05}
obtained in lattice gauge calculations as the color-singlet $Q$-$\bar
Q$ potential without rigorous theoretical justifications.  The
internal energy $U_1$ is significantly deeper and spatially more
extended than the free energy $F_1$. Conclusions will be quite
different if one uses the free energy $F_1$ or the linear combination
of $F_1$ and $U_1$ defined by Eq.\ (\ref{Uqq1}) below as the $Q$-$\bar
Q$ potential.

If one constructs the Schr\" odinger equation for the color-singlet
$Q$ and $\bar Q$, the $Q$-$\bar Q$ potential $U_{Q\bar Q}^{(1)}({\bf
  r},T)$ in the Schr\" odinger equation contains those interactions
that act on $Q$ and $\bar Q$, when the $Q$ and $\bar Q$ are separated
by ${\bf r}$ at temperature $T$ and the medium particles have
re-arranged themselves self-consistently.  As shown theoretically in
detail in \cite{Won05} for lattice gauge theory, this potential is
given by
\begin{eqnarray}
\label{eqU}
U_{Q\bar Q}^{(1)}({\bf r},T)= U_1({\bf r},T)-[ U_g({\bf r},T)-U_{g0}],
\end{eqnarray}
where $U_1({\bf r},T)$ is the color-singlet internal energy, $U_g({\bf
  r},T)$ and $U_{g0}$ are gluon internal energy in the presence and
absence of the color-singlet $Q$ and $\bar Q$ pair, respectively.
This is a rather general result when screening occurs, as a similar
relationship exists between the total internal energy and the
$Q$-$\bar Q$ potential in the analogous case of Debye screening
\cite{Won06}.

We proposed earlier a method to determine the gluon energy in
Eq. (\ref{eqU}) in terms of the gluon entropy by making use of the
equation of state of the quark-gluon plasma obtained in an independent
lattice gauge calculation in quenched QCD \cite{Won05,Won05a,Won06}.
The equation of state of the medium provides a relationship between
the QGP internal energy density $\epsilon_g$ and the QGP entropy
density $\sigma$,
\begin{eqnarray}
\epsilon_g=T\sigma -p.
\end{eqnarray}
Thus, by expressing $p$ as $(3p/\epsilon_g) (\epsilon_g/3)$ with the
ratio $a(T)=3p/\epsilon_g$ given by the known equation of state in
quenched QCD, the plasma internal energy density $\epsilon_g$ in
quenched QCD is related to the entropy density $T\sigma$ by
\begin{eqnarray}
\epsilon_g=\frac{3}{3+a(T)} T\sigma .
\end{eqnarray} 
This is just
\begin{eqnarray}
\epsilon_g=\frac{dU_g^{(1)}}{dV} =\frac{3}{3+a(T)}\frac{d}{dV}\int dV~
T(\sigma-\sigma_0+\sigma_0),
\end{eqnarray}
where $\sigma_0$ is the entropy density in the absence of $Q$ and
$\bar Q$.  Noting that the entropy of the medium for the color-singlet
$Q$-$\bar Q$ pair is $TS_1=\int dV~ T(\sigma-\sigma_0)$ and
$U_{g0}$ is related to $\int dV~T\sigma_0$, the above equation
leads to
\begin{eqnarray}
\frac{d[U_g^{(1)}({\bf r},T)-U_{g0}(T)]}{dV} =\frac{3}{3+a(T)}
\frac{T\,dS_1({\bf r},T)}{dV}.
\end{eqnarray} 
The plasma internal energy difference integrated over the volume is
therefore given by
\begin{eqnarray}
U_g^{(1)}({\bf r},T)-U_{g0}(T)=\frac{3}{3+a(T)}TS_1({\bf r},T).
\end{eqnarray}
But $TS_1({\bf r},T)$ has already been obtained as $TS_1({\bf
r},T)=U_1({\bf r},T)-F_1({\bf r},T)$ from the lattice gauge
calculations \cite{Kac03}.  The plasma internal energy difference,
$U_g^{(1)}({\bf r},T)-U_{g0}$, is therefore equal to
\begin{eqnarray} 
U_g^{(1)}({\bf r},T)-U_{g0}=\frac{3}{3+a(T)}[U_1({\bf r},T)-F_1({\bf r},T)].
\end{eqnarray}
The $Q$-$\bar Q$ potential, $U_{Q\bar Q}^{(1)}({\bf r},T)$, as
determined from Eq.\ (\ref{eqU}) by subtracting the above plasma
internal energy difference from $U_1({\bf r},T)$, is then a linear
combination of $F_1({\bf r},T)$ and $U_1({\bf r},T)$ given by
\cite{Won05}
\begin{eqnarray}
U_{Q\bar Q}^{(1)}({ \bf r},T)= 
 \frac{3}{3+a(T)}    F_1({ \bf r},T) 
+\frac{a(T)}{3+a(T)} U_1({ \bf r},T).
\end{eqnarray}
We can rewrite the above as 
\begin{eqnarray}
\label{Uqq1}
W_1({\bf r},T))\equiv U_{Q\bar Q}^{(1)}({\bf r},T)= 
 f_F(T)    F_1({\bf r},T) 
+f_U(T)  U_1({\bf r},T),
\end{eqnarray}
where for brevity of notation we have renamed $U_{Q\bar
  Q}^{(1)}({\bf r},T)$ as  $W_1({\bf r},T)$
and we have defined the coefficients
\begin{eqnarray}
f_F(T)=\frac{3}{3+a(T)},
\end{eqnarray}
and
\begin{eqnarray}
f_U(T)=\frac{a(T)}{3+a(T)}.
\end{eqnarray}
To determine $a(T)$, we use the equation of state of Boyd $et~al.$
\cite{Boy96} for quenched QCD. The values of $a(T)$, $f_F (T)$, and
$f_U(T)$ are given in Fig.\ (1) of Ref.\ \cite{Won05}.  The $W_1({\bf
  r},T)$ potential is approximately $F_1({\bf r},T)$ near $T_c$ and is
$3F_1({\bf r},T)/4+U_1({\bf r},T)/4$ for high temperatures
\cite{Won05,Won05a,Won06}.

To evaluate the $Q$-$\bar Q$ potential $W_1({\bf r},T)$, we use the
free energy $F_1({\bf r},T)$ and the internal energy $U_1({\bf r},T)$
obtained by Kaczmarek $et~al.$ in quenched QCD \cite{Kac03} for which
$F_1({\bf r},T)$ and $U_1({\bf r},T)$ can be parametrized in terms of
a screened Coulomb potential,
\begin{eqnarray}
\label{f1}
F_1( {\bf r},T)=C_F(T)
-\frac{4}{3}\frac{\alpha_{F}(T)e^{-\mu_F (T) r}}{r},
\end{eqnarray}
and 
\begin{eqnarray}
\label{u1}
U_1({\bf r},T)=C_U(T)
-\frac{4}{3}\frac{\alpha_Ue^{-\mu_U(T) r}}{r}.
\end{eqnarray}
The parameters $C_{F,U}(T)$, $\alpha_{F,U}(T)$, and $\mu_{F,U}(T)$
are shown in Figs. 2 and 3 of Ref.\ \cite{Won05}.

The non-relativistic Schr\" odinger equation for the $Q\bar Q$ states
in the $W_1({\bf r},T)$ potential is
\begin{eqnarray}
  \left \{ \frac{ {\bf p}^2 } {2\mu_m} + W_1({\bf r},T) + m_Q + m_{\bar Q} \right \}
  \psi({\bf r},T) = M(T) \psi({\bf r},T).    
\end{eqnarray}
This equation can be re-written as
\begin{eqnarray}
\left \{ \frac{{\bf p}^2}{2\mu_m} + W_1({\bf r},T)- W_1({\bf r}\to \infty,T)
\right \} \psi({\bf r},T) &=& \{ M(T)  -   m_Q - m_{\bar Q} -  W_1({\bf r}\to \infty,T) \} \psi({\bf r},T)\nonumber\\
&\equiv& \epsilon(T) \psi({\bf r},T).    
\end{eqnarray}
With the $W_1({\bf r},T)$ potential as given by Eq.\ (\ref{Uqq1}),
(\ref{f1}), and (\ref{u1}), the mass of the composite meson and the
eigenenergy $\epsilon(T)$ of the above equation are related by
\begin{eqnarray}
\label{mass}
M(T)= m_Q + m_{\bar Q} +\epsilon(T) +C(T), 
\end{eqnarray}
where $C(T)$ is the asymptotic value of the $W_1({\bf r}\to \infty,T)
[\equiv U_{Q\bar Q}^{(1)}({\bf r}\to \infty,T)]$ obtained in lattice
gauge calculations,
\begin{eqnarray}
C(T) = W_1({\bf r}\to \infty,T)=
 f_F(T) C_F(T) + f_U(T) C_U(T).
\end{eqnarray}

For simplicity, we ignore spin and we study the $S$-wave charmonium
state which splits into $J/\psi$ and $\eta_c$ when the spin-spin
interaction is included.  Using the $W_1({\bf r},T)$ potential and
$m_c=1.41$GeV, we calculate the bound $L$=0 charmonium state energy
$\epsilon(T)$.  The mass $M(T)$ of the bound $S$ state can then be
obtained from $\epsilon(T)$ using Eq.\ (\ref{mass}) and $C(T)=W_1({\bf
r}\to \infty,T)$.  We list $C(T)$, $\epsilon(T)$, and $M(T)$ for the
$L$=0 charmonium states in Table I.  We also list $M(T=0)=3.064$ GeV
given by $M(S{\rm ~state},T=0)=3M(J/\psi,T=0)/4+M(\eta_c,T=0)/4$ where
$ M(J/\psi,T=0)$ and $M(\eta_c,T=0)$ are experimental masses.

\vskip 0.4cm \centerline{Table I.  The quantities $C(T)$,
$\epsilon(T)$, $M(T)$ (all in GeV) for the bound $L$=0 charmonium
state, } \centerline{ calculated with the $W_1({\bf r},T)$ potential
of Eq.\ (\ref{Uqq1}) in quenched QCD.  }

{\vskip 0.3cm\hskip 1.4cm
\begin{tabular}{|c|c|c|c|c|c|c|c|c|}
\hline
  {$T$}                         &  0.           & 1.13$T_c$  &  1.18$T_c$ 
                                & 1.25$T_c$  &  1.4$T_c$  &  1.6$T_c$ 
                                & 1.95$T_c$  &  2.60$T_c$ \\
\hline
$C(T)$                             &           &  0.2962    &  0.2710
                                 &   0.2476 &   0.2218   &   0.1991  
                                 &   0.1620 & 0.1094 \\
\hline
Bound state energy $\epsilon(T)$      &           &   -0.0340 &  -0.02078
                                 &   -0.0105 &  -0.0036   &  -0.00019
                                 & unbound & unbound   \\
\cline{1-7}
Mass $M(T)$                  &   3.064  &  3.082     &  3.070    
                                 &   3.057  &  3.038     &  3.019 
                                 &   &   \\
\hline
\end{tabular}
}
\vspace*{0.3cm}

As one observes, the bound $S$ state masses obtained in the potential
model with the potential of Eq.\ (\ref{Uqq1}) are nearly the same as
the $S$-wave charmonium mass at $T=0$.  They change only very slightly
with temperature until the $S$-wave charmonium dissociates at
$\sim$1.6$T_c$.  The dominant errors in the mass value $M(T)$ comes
from the statistical errors in the asymptotic values $C_{F,U}(T)$ of
$F_1({\bf r},T)$ and $U_1({\bf r},T)$ in \cite{Kac03}.  They leads to
uncertainties of about $\pm 0.7\%$ in the values of $M(T)$.
 
\vskip 0.4cm \centerline{Table II.  Spontaneous dissociation
  temperatures calculated with the $W_1({\bf r},T)$ potential of Eq.\
  (\ref{Uqq1}), } \centerline{ the $F_1({\bf r},T)$ potential, and the
  $U_1({\bf r},T)$ potential, in quenched QCD.}

 {\vskip 0.3cm\hskip 2.5cm
\begin{tabular}{|c|c|c|c|c|c|c|c|}
\hline
     & \multicolumn{4}{c|}{Dissociation temperatures in Quenched QCD}      \\    \cline{1-5}
\hline
{\rm States     } & {\rm Spectral Analysis} 
                 & $W_1$ Potential        
                 & $F_1$ Potential        
                 & $U_1$ Potential        
\\
\hline
$J/\psi,\eta_c$       &   $1.62$-$1.70T_c^\dagger$ 
&   $1.62\,T_c$  
&   $1.40 \,T_c$ & $2.60 \,T_c$ 
\\ \cline{1-5}  
$\chi_c$              & below $ 1.1T_c^{\natural}$   
& unbound in QGP
& unbound in QGP            &  $1.18 \,T_c$ 
\\ \cline{1-5}
$\psi',\eta_c'$       &  
&   unbound in QGP
&   unbound in QGP & $1.23 \,T_c$ 
\\ \cline{1-5}  
$\Upsilon,\eta_b$     &                   
&  $ 4.1 \,T_c$    
&    $ 3.5 \,T_c$   &    $ \sim 5.0  \,T_c$ 
\\ \cline{1-5}  
$\chi_b$              &  $1.15$-$1.54T_c^{\sharp}$     
&  $ 1.18 \,T_c$      
&  $ 1.10 \,T_c$      & $ 1.73 \,T_c$  
\\ \cline{1-5}
$\Upsilon',\eta_b'$     &                   
&  $ 1.38 \,T_c$    
&    $ 1.19  \,T_c$   &    $ 2.28  \,T_c$  
\\ \cline{1-5}
                                                        \hline
\multicolumn{5}{l}
{${}^\dagger$Ref.\cite{Asa03},~~~     
 ${}^{\natural}$Ref.\cite{Dat03},~~~     
 ${}^{\sharp }$Ref.\cite{Pet05,Jak06}}     
\end{tabular}
}
\vspace*{0.3cm}

In lattice gauge spectral function analyses, the positions of the
bound $S$ states do not appear to change significantly up to $1.5T_c$
\cite{Asa03,Mat01,Dat03,Pet05,Jak06} in agreement with the small
variation of $M(T)$ in the potential model using the $W_1({\bf r},T)$
potential as shown in Table I.  The widths of many color-singlet heavy
quarkonia are broadened suddenly at various temperatures
\cite{Asa03,Mat01,Dat03,Pet05,Jak06}.  From the shape of the spectral
functions, the range of temperatures from 1.62$T_c$ to $1.70T_c$, in
which the $J/\psi$ width is broadened suddenly, corresponds to the
range of $J/\psi$ spontaneous dissociation temperatures.  Dissociation
temperatures for $\chi_c$ and $\chi_b$ in spectral analyses in
quenched QCD have also been obtained \cite{Dat03,Pet05,Jak06}.
Dissociation temperatures obtained with the $W_1({\bf r},T)$ potential
of Eq.\ (\ref{Uqq1}) as well as the $F_1({\bf r},T)$ and $U_1({\bf
r},T)$ potentials are given in Table II.  The dissociation
temperatures obtained in the $W_1({\bf r},T)$ potential are found to
give the best agreement with the dissociation temperatures obtained in
lattice gauge spectral function analyses, as shown in Table II
\cite{Won05,Won05a}.  It is therefore of great interest to see whether
the correlators obtained from such a potential agree with those from
lattice gauge calculations.

\section{Evaluation of the Meson Correlator in the Potential Model}

Upon limiting our attention to the pseudoscalar charmonium with $L$=0, we
note that the quark-antiquark potential itself does not possess a
potential barrier, and thus there are no $S$-wave resonances
\cite{New66}.  We calculate the $S$-wave phase shift using the
amplitude-phase method of Wheeler \cite{Whe37} and Calogero
\cite{Cal67,Won97}.  The $S$-wave phase shifts as a function of the
continuum energy $\epsilon$ and temperature $T/T_c$ are shown in
Fig. 2.  We note that the phase shifts behave in two different ways
depending on whether there are bound $S$ states.

\begin{figure} [h]
\includegraphics[angle=0,scale=0.50]{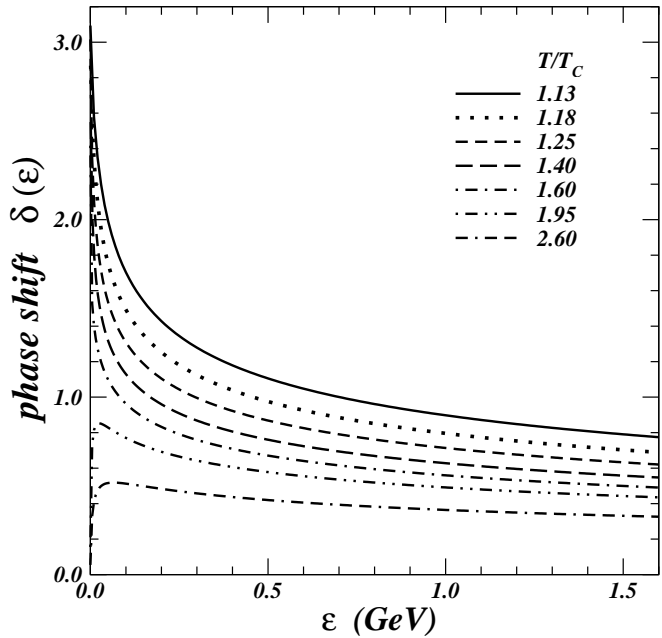}
\caption{ Phase shift as a function of the composite continuum energy
  $\epsilon$ and the plasma temperature calculated in the potential
  model using the $W_1({\bf r},T)$ potential.  }
\end{figure}

When there are $m$ bound $S$ states in the potential, the $S$ wave
phase shift at zero energy will start at $m\pi$, in accordance with
the Levinson theorem.  For the $W_1({\bf r},T)$ potential, there is
one bound S-wave state for temperatures below $\sim$1.6$T_c$, as
indicated by the phase shift of $\delta(\epsilon=0)$=$\pi$.  The phase
shift gradually decreases as energy increases and $\partial
\delta(\epsilon)/\partial \epsilon$ is negative for all continuum
energies.  In this case with one or many $S$-wave bound states, the
time delays for all continuum states are negative and there are no
Gamow states in the continuum.  Thus, in Eq.\ (\ref{Gtau}), only a
single term, a bound state term with a bound state mass $M_b$,
contributes to the correlator $G(\tau)$ for $T< \sim 1.6 T_c$.  The
correlator $G(\tau)$ of Eq.\ (\ref{Gtau}) becomes
\begin{eqnarray}
G(\tau)\propto 
\frac {\cosh\{M_b(T)(\tau-\beta/2)\}}
{\sinh \{M_b(T)\beta/2)\}}
\end{eqnarray}
where $M_b(T)$ is the mass of the bound $S$ state at the temperature
$T$.  A lattice gauge correlator $G(\tau)$ is represented in terms of
its ratio with respect to a ``reconstructed'' correlator $G_{\rm
recon}(\tau)$, which is defined as the correlator calculated with the
``reconstructed'' spectrum at $T=0$.  In actual practice in the
evaluation of the lattice gauge correlators $G_{\rm recon}(\tau)$ to
obtain the ratios of $G(\tau)/G_{\rm recon}(\tau)$ shown in Figs.\ 1a
and 3a, Ref.\ \cite{Dat03} has used the ``reconstructed'' spectrum at
$T=0.75T_c$ which contains only a single bound state \cite{footnote2}.
Therefore, in order to compare with lattice gauge $G(\tau)/G_{\rm
recon}(\tau)$, we need to include only a single lowest mass bound
state in Eq.\ (\ref{Gtau}) to evaluate the ``reconstructed''
correlator $G_{\rm recon}(\tau)$ in the potential model.  The
potential model correlator $G(\tau)$ relative to the potential model
$G_{\rm recon}(\tau)$, normalized to unity at $\tau=0$, is thus
\begin{eqnarray}
\label{GG}
\frac{G(\tau)}{G_{\rm recon}(\tau)} =  
a \frac {\cosh\{M_b(T)(\tau-\beta/2)\}}
{\sinh \{M_b(T)\beta/2)\}}
\Biggr /
\frac {\cosh\{M_b(T=0)(\tau-\beta/2)\}}
{\sinh \{M_b(T=0)\beta/2)\}},
\end{eqnarray}
where $a=\tanh(M_b(T)\beta/2)/\tanh(M_b(T=0)\beta/2) \sim 1$.  

\null{}
\vspace*{-0.0cm}
\begin{figure} [h]
\includegraphics[angle=0,scale=0.80]{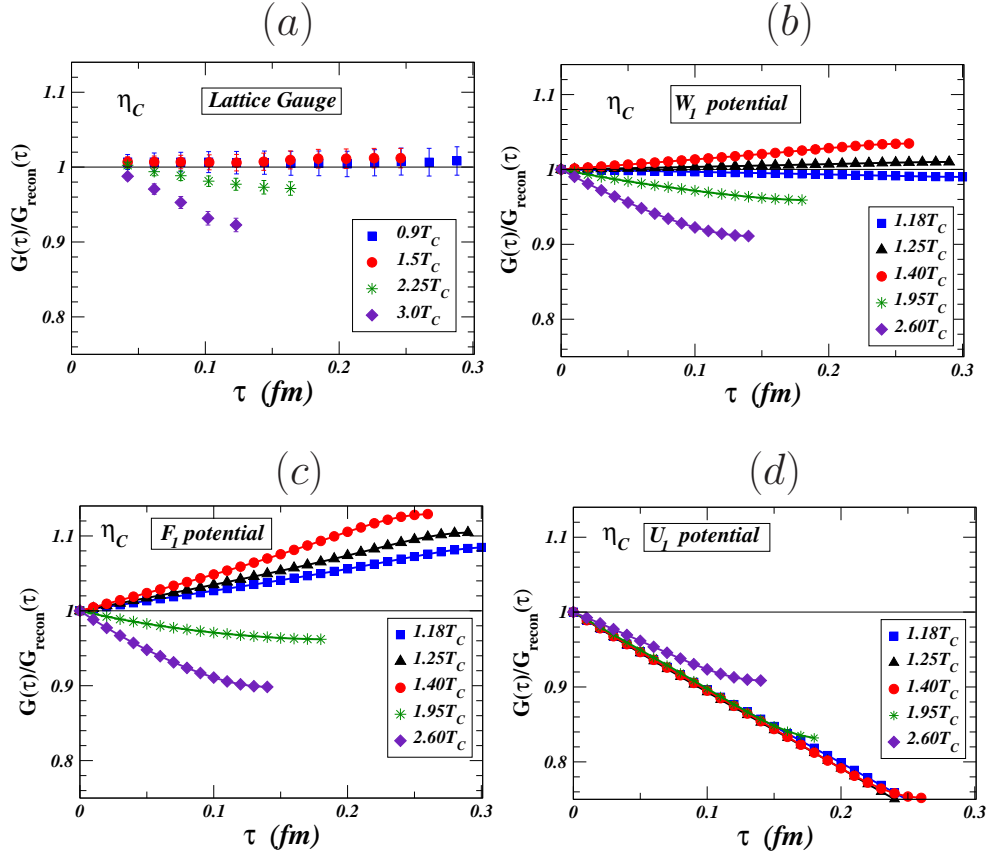}

\vspace*{-9.50cm}
\caption{ Fig. 3a shows the pseudoscalar charmonium correlators
  $G(\tau)/G_{\rm recon}(\tau)$ obtained in lattice gauge calculations
  \cite{Dat03}. The potential model pseudoscalar charmonium
  correlators $G(\tau)/G_{\rm recon}(\tau)$ obtained with the
  $W_1({\bf r},T)$ potential of Eq.\ (\ref{Uqq1}) are shown in Fig.
  3b, with the free energy $F_1({\bf r},T)$ potential in
  Fig. 3c, and with the internal energy energy $U_1({\bf r},T)$
  potential in Fig.\  3d.  }
\end{figure}

Using the above equation that is based on the concept of the absence
of $L$=0 resonance states and Gamow states when a bound state is present, we
calculate the potential model correlators $G(\tau)/G_{\rm
recon}(\tau)$ for the cases of 1.18$T_c$, 1.25$T_c$, and 1.40$T_c$
with bound state masses $M(T)$ obtained in the $W_1({\bf r},T)$
potential as given in Table I.  The results for these three
temperatures are shown in Fig.\ 3b.  As the bound state mass values in
Table I have uncertainties of about $\pm$0.7\% in $M(T)$, the
uncertainties of the potential model correlators are about $\pm 0.007$
in $G/G_{\rm recon}$.  The potential model correlators can be compared
with the correlators obtained in lattice gauge calculations show in
Fig. 3a \cite{Dat03}.  The general features of the potential model
correlators at these three temperatures below about 1.6$T_c$ agree
with those of the lattice gauge correlators.  In particular, the
slopes $d[G/G_{\rm recon}]/d\tau$ in Fig.\ 3b for $T<\sim 1.6 T_c$ in
potential model calculations are of the order $\sim 0.03-0.1/$fm,
which do not differ much from the nearly zero slopes of $d[G/G_{\rm
recon}]/d\tau$ in Fig.\ 3a for $T=0.9T_c$ and 1.5$T_c$ in lattice
gauge calculations.  They differ significantly from the general shapes
of the potential model correlators obtained by the authors of Ref.\
\cite{Moc06,Moc06a} in Figs.\ 1b and 1c, where $d[G/G_{\rm
recon}]/d\tau \sim 1/$fm in Fig.\ 1b and $\sim 2.5/$fm in Fig.\ 1c,
for $\tau<0.2$fm.

As the temperature increases above $\sim$1.6$T_c$, the $S$-wave state
calculated with the $W_1({\bf r},T)$ potential is no longer bound and
the phase shifts at $T/T_c=1.95$ and $2.60$ are shown in Fig. 2.  For
these cases without a bound $S$-wave state, the phase shift starts at
zero at zero continuum energy $\epsilon$ and it increases rapidly as
the energy increases.  After reaching a peak value below $\pi/2$, the
phase shift decreases slowly as the energy increases.  Thus, there is
a region of continuum states for which $\partial
\delta(\epsilon)/\partial \epsilon$ is positive.  They possess
positive time delays and lifetimes.  They are Gamow states capable of
propagating as a composite meson to contribute to the meson
correlator.  In this case without an $S$-wave bound state, there are
no contributions from bound states to the correlator $G(\tau)$.  Only
continuum Gamow states represented by the integral over ${\bf k}$
contribute to the meson correlator $G(\tau)$ in Eq.\ (\ref{Gtau}).

\begin{figure} [h]
\hspace*{1.3cm}
\includegraphics[angle=0,scale=0.50]{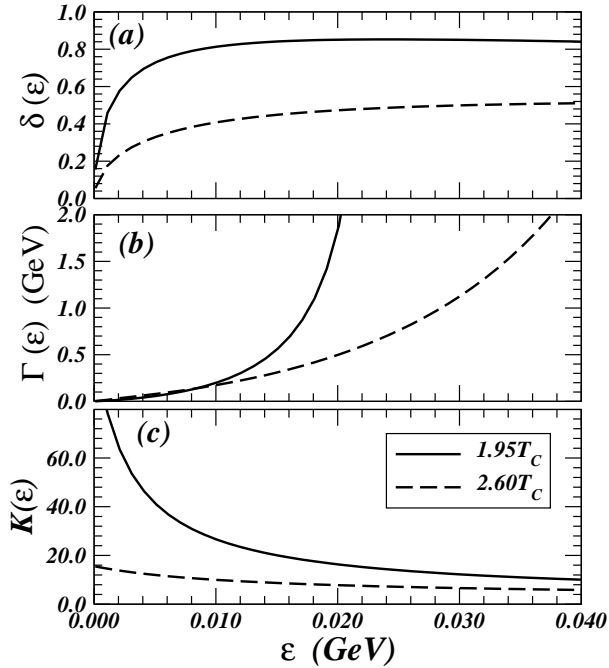}
\caption{ (Fig.\ 4a) The phase shift $\delta(\epsilon)$ as a function
  of the continuum energy $\epsilon$ at $T=1.95T_c$ and $2.6T_c$,
  calculated in the $W_1({\bf r},T)$ potential of Eq.\ (\ref{Uqq1}).
  (Fig.\ 4b) The width of the Gamow states $\Gamma(\epsilon)$ as a
  function of the continuum energy $\epsilon$.  (Fig. 4$c$) ${\cal
  K}(\epsilon)$ as a function of $\epsilon$.  }
\end{figure}

An expanded view of the phase shift $\delta(\epsilon)$ as a function
of the continuum state ${\bf k}$ with energy $\epsilon={\bf
k}^2/2\mu_m$ is shown in Fig. 4a for $T=1.95 T_c$ and $T=2.60 T_c$.
The corresponding width $\Gamma(\epsilon)$ extracted from the time
delay $\partial \delta(\epsilon)/\partial \epsilon)$ is shown in 4b.
At $T=1.95 T_c$, the width is zero at $\epsilon$=0, and the gradual
increase in width turns into a rapid increase as $\epsilon$ increases.
The increase in width is less rapid at the higher temperature of
2.6$T_c$.  Above the energy $\epsilon({\rm max})$ at which
$\Gamma(\epsilon({\rm max}))=\Gamma_{\rm max}$ where $\Gamma_{\rm
max}=2T$, the width of the continuum state is either too large or
$\partial \delta(\epsilon)/\partial \epsilon$ is negative, and
continuum states with energy above $\epsilon({\rm max})$ will not
contribute to the meson correlator in the integral over ${\bf k}$.
For $T=1.95T_c$, $\epsilon({\rm max})$ is 0.018GeV, and for
$T=2.60T_c$, $\epsilon({\rm max})$ is 0.033GeV. They are located at
energies slightly above the continuum threshold.

We can also evaluate the amplitude of the continuum wave function at
the origin.  The comparison of the absolute square of the amplitude at
the origin ${\bf r}=0$, with the absolute square of the amplitude at
very large ${\bf r}$ gives the ${\cal K}$-factor for the continuum
state \cite{Won97} shown in Fig.\ 4c.  This ${\cal K}$-factor is quite
large at low energies, due to the attractive interaction between the
quark and the antiquark.  The ${\cal K}$-factor decreases as a
function of the continuum energy $\epsilon$.  We need to specify the
width limit $\Gamma_{\rm max}(T)=1/\tau_{\rm min~life}(T)=2T$ by using
Eq.\ (\ref{gammax}).  For $T=1.95T_c$, $\tau_{\rm min~life}$ is
0.185fm and $\Gamma_{\rm max}$ is 1.05GeV.  For $T=2.6T_c$, $\tau_{\rm
min~life}$ is $0.141$fm and $\Gamma_{\rm max}$ is 1.40GeV.

After setting the limit on the width of the Gamow states, we can
calculate the pseudoscalar meson correlator, $G(\tau)/G_{\rm
recon}(\tau)$ normalized to unity at $\tau=0$, for the case without a
bound state.  The results of the potential model correlators for $1.95
T_c$ and $2.6 T_c$ using the $W_1({\bf r},T)$ potential are shown in
Fig.\ 3b.  As one observes, the general trends of the correlators at
these temperatures agree with those from the lattice gauge
correlators.

The results in Figs.\ 3a and 3b indicate that when the contributions
from bound states and continuum states are properly treated, there is
indeed agreement between the lattice gauge correlators and the
potential model correlators using the $W_1({\bf r},T)$ potential that
is a linear combination of $F_1({\bf r},T)$ and $U_1({\bf r},T)$.

It is of great interest to investigate whether the potential model
correlators depends on the potential.  Accordingly, we evaluate the
potential model correlators using the color-singlet free energy
$F_1({\bf r},T)$ \cite{Dig01a,Won01a,Bla05} and the internal energy
$U_1({\bf r},T)$ \cite{Kac02,Shu03,Alb05,Man05} for comparison.  We
need the bound state masses at various temperatures to evaluate the
correlators, as required by Eqs.\ (\ref{Gtau}) and (\ref{GG}).  We
list state energy $\epsilon(T)$ and bound state mass $M(T)$ calculated
in the $F_1({\bf r},T)$ potential and the $U_1({\bf r},T)$ potential
for the $L$=0 charmonium in Table III.

\vskip 0.4cm \centerline{Table III.  Bound state energy $\epsilon(T)$
  and bound state mass $M(T)$ (all in GeV) for the $L$=0 charmonium }
  \centerline{ calculated with the $F_1$ and
  $U_1$ potentials in quenched QCD.  }

{\vskip 0.3cm\hskip 1.4cm
\begin{tabular}{|c|c|c|c|c|c|c|c|c|}
\hline
  \multicolumn{2}{|c|} {$T$}                  & 1.13$T_c$  &  1.18$T_c$ 
                                & 1.25$T_c$  &  1.4$T_c$  &  1.6$T_c$ 
                                & 1.95$T_c$  &  2.60$T_c$ \\
\hline
$F_1$ Potential &                    $\epsilon(T)$   &   -0.0117 &  -0.0051
                                 &   -0.0018 &  -0.000007   &  unbound
                                 & unbound & unbound   \\
\cline{2-6}
   & Mass $M(T)$                 &   3.029  &  3.012     &  2.997    
                                 &   2.972  &            &  
                                 &            \\
\hline
$U_1$ Potential &                    $\epsilon(T)$   &   -0.7483 &  -0.4422
                                 &   -0.2613 &  -0.1331   &  -0.07820
                                 & -0.0669 & -0.0225   \\
 \cline{2-9}
   & Mass $M(T)$                 &   3.188  &  3.279     &  3.285    
                                 &   3.279  &  3.265     &  3.259 
                                 &   3.201         \\
\hline
\end{tabular}
}
\vspace*{0.3cm}

We note that as a function of temperature $T$, the $L=0$ charmonium masses
with the $F_1$ potential are lower than the mass at $T=0$ and
they decrease slowly with temperature up to its dissociation temperature
at $\sim 1.4T_c$.  On the other hand, for the $U_1$ potential the masses
are greater than the charmonium mass at $T=0$ and vary only slightly  with
temperature.

We need to calculate further the phase shifts and the wave function
amplitude at the origin and evaluate the correlators in these
potential models.  The potential model correlators obtained with $F_1$
and $U_1$ as the potentials are shown in Fig. 3c and 3d respectively.
The comparison of these potential model correlators with the lattice
gauge correlators indicates that the correlators calculated with the
$F_1({\bf r},T)$ and $U_1({\bf r},T)$ potentials deviate from the
lattice gauge correlators, the deviation being greater for the $U_1$
potential than the $F_1$ potential.  The comparison shows that among
the three different potentials, correlators obtained with
the $W_1({\bf r},T)$ potential  of \cite{Won05,Won05a,Won06,Won06a} give
the best agreement with the lattice gauge correlators.

\section{Implications on the Lattice Gauge Spectral Function Analysis }

The analysis of spectral function at finite temperatures in lattice
gauge theory contains systematic uncertainties and lattice artifacts.
In the maximum entropy method used for the spectral function analysis,
it is necessary to assume a default spectrum to define the entropy.
The spectral function obtained in the maximum entropy method depends
on the assumed default spectrum (see Fig.\ 4 of \cite{Dat03}).  The
extracted spectra often exhibit two prominent broad peaks in the
continuum which do not seem to correspond to physical continuum states
\cite{Asa03,Mat01,Dat03,Pet05,Jak06}.  While it is important to study
theoretically what the lattice artifacts are expected to be, some
prior knowledge of the spectral function as inferred from the
potential model will be useful to provide additional information for
the assumed default spectrum and the expected spectral function.

From the physical picture that emerges in the evaluation of the
current-current correlator, the spectral function analysis in the
lattice gauge theory should be guided by the basic principle that out
of the set of continuum states, only resonance states and Gamow states
with lifetimes of sufficient magnitudes can propagate as composite
objects and can contribute to the current-current correlator.  The
potential model can provide useful information on the nature of bound
states, resonance states, and Gamow states for spectral function
analysis.  We can discuss first the case of $L=0$.  As the $Q$-$\bar
Q$ interaction itself does not possess a barrier, we expect that there
are no $L=0$ potential resonances in the continuum \cite{New66}.  The
occurrence of $L=0$ bound states in the potential is accompanied by
phase shifts with negative $\partial \delta (\epsilon)/\partial
\epsilon$ (Fig.\ 2) with no delay time and composite lifetime for the
continuum states.  Consequently when $L=0$ bound states occur, only
bound states can contribute to the correlator.  According to these
results of the potential model, the lattice gauge spectral function
analysis for $L=0$ states should include only bound states without
continuum states. On the other hand, when the $L=0$ bound states are
absent, states with delay time and widths occurs at energies close to
$\epsilon\sim 0$.  The correlator receives contribution only from this
region of continuum states (see Fig. 3b) and lattice gauge spectrum
function analysis for $L=0$ states should employ a default spectrum
that contains this element of the continuum contribution.

For the case of $L>0$, the presence of centrifugal barrier leads to
potential pockets and possible potential resonance states, which may
or may not occur, depending on the strength of the potential and the
centrifugal barrier.  One expects that when potential resonance states
occur, the phase shift decreases as a function of $\epsilon$ at
continuum energies above the resonance energies, and only bound states
and possible resonance states contribute to the correlator in lattice
gauge analysis.  On the other hand, when bound states and resonance
states are absent (as for example at high temperatures), Gamow states
with delay time and widths of various magnitude occurs.  The
correlator receives contribution only from these Gamow states.
Lattice gauge spectrum function analysis for $L>0$ states should
employ a default spectrum that contains these essential
characteristics.

\section{Conclusions and Discussions}

The potential model has been a useful concept in the physics of heavy
quarkonia since the discovery of $J/\psi$.  It provides a useful tool
to examine quarkonium energies, quarkonium wave functions, reaction
rates, transition rates, and decay widths
\cite{God85,Bar92,Won02,Cra04,Cra06}. It allows the extrapolation to
the region of high temperatures by expressing screening effects in
terms of the temperature dependence of the potential
\cite{Mat86,Dig01a,Won01a,Kac02,Shu03,Won05,Won05a,Won06,Won06a,Son05,Alb05,Bla05,Man05,Dig05}.
The comparisons of the potential model dissociation temperatures with
the lattice gauge spectral function dissociation temperatures show
consistency when one uses the potential that is a linear combination
of the free energy and the internal energy proposed in
\cite{Won05,Won05a}.

In addition to lattice gauge spectral function analyses, results have
also been obtained for the lattice gauge current-current correlators
\cite{Moc06,Moc06a}.  The current-current correlator is related to the
spectral function by a generalized Laplace transform.  In principle,
they carry equivalent information. One expects that the consistency of
the potential model with lattice gauge spectral function analyses
should be extended to the comparison of the potential model
current-current correlators with lattice gauge current-current
correlators.  Recent works in Refs.\ \cite{Moc06,Moc06a} however make
the contrary claim that the meson correlators obtained from many
different types of potential models are not consistent with lattice
gauge correlators and potential models cannot describe heavy quarkonia
above $T_c$. In the work of Ref.\ \cite{Moc06,Moc06a}, continuum
states arising from a free fermion $Q$ and $\bar Q$ pair in a fermion
gas contribute to the correlator.  However, to be able to propagate as
a composite meson so as to contribute to the correlator, the quark and
antiquark must be temporally and spatially correlated.  Continuum
states in the free quark and antiquark gas may not have sufficient
temporal and spatial correlations to qualify as composite objects
for such a propagation.

Following the basic principle that among the continuum states only
resonance states and Gamow states with a lifetime of sufficient
magnitude can propagate as a composite object and contribute to the
correlator, we re-evaluate the current-current correlator in the
potential model.  With the simple example of the pseudoscalar
correlator, we show in this paper that when the bound state and
continuum state contributions are properly treated, the $W_1({\bf
r},T)$ potential using a linear combination of $F_1$ and $U_1$
proposed in \cite{Won05,Won05a} gives correlators consistent with
those of lattice gauge correlators, while the $F_1$ potential and the
$U_1$ potential lead to deviations.  Our results indicate consistency
of the $W_1({\bf r},T)$ potential proposed in \cite{Won05,Won05a} with
both the lattice gauge spectral function analysis and the lattice
gauge correlator analysis.  The present agreement is not surprising as
the current-current correlator and the spectral function are related
by a generalized Laplace transform, and they indeed carry equivalent
information.

There are uncertainties, limitations, and future prospects in the
potential model that requires further investigations.  Lattice gauge
calculations provide information on thermodynamical quantities of the
free energy $F_1$ and the internal energy $U_1$.  We have shown
deductively in Eq.\ (11) of Ref.\ \cite{Won05} that the internal
energy $U_1({\bf r},T)$ contains $U_{Q\bar Q}^{(1)}({\bf r},T)$ and
$U_g^{(1)}({\bf r},T)-U_{g0}(T)$.  Only $U_{Q\bar Q}^{(1)}({\bf r},T)$
pertains to the interaction potential between $Q$ and $\bar Q$.  The
other parts need to be subtracted out from $U_1({\bf r},T)$ to obtain
the $Q$-$\bar Q$ potential.  We have made use of the knowledge of the
equation of state from an independent lattice gauge calculations to
evaluate $U_{Q\bar Q}^{(1)}({\bf r},T)$ from $U_1({\bf r},T)$, leading
to the present linear-combination model of the $U_{Q\bar Q}^{(1)}({\bf
r},T)$ potential of Eq.\ (\ref{Uqq1}).  However, a more rigorous
treatment will involve a direct evaluation of the $Q$-$\bar Q$
potential by evaluating the quantities of $U_g^{(1)}({\bf
r},T)-U_{g0}(T)$ in the lattice gauge theory.  It will be of great
interest to examine how one can determine directly the $Q$-$\bar Q$
potential in a lattice gauge calculations without resort to the
equation of state.

The potential model we have developed so far has the limitation that
the important spin-spin interaction has not been included.  As the
spin-spin interaction is responsible for the $\eta_c$-$J/\psi$ and
$\eta_b$-$\Upsilon$ splittings, it modifies the binding energies and
the dissociation temperatures.  It is important to include spin-spin
and other components of the $Q$-$\bar Q$ interactions in the potential
model to see how they may affect the stability of heavy quarkonia.

The potential model allows the evaluation of many quantities of
interest.  The potential model in the present manuscript uses
thermodynamical quantities obtained in quenched QCD.  The effects of
dynamical quarks on the stability of quarkonia can be studied by using
potential models extracted from thermodynamic quantities calculated in
full QCD \cite{Kac05,Won05a}.  One can examine the quark mass
dependence of quarkonium stability and quarkonium gluon dissociation
cross sections, some results of which have been presented recently
\cite{Won06,Won06a}.  The heavy quark potential model so far developed
has the limitation that it is restricted to non-relativistic cases.
It will also be of interest to study the relativistic effects by
examining relativistic two-body potential models along the lines of
Dirac's constraint dynamics as in Ref.\ {\cite{Cra04,Cra06}, which
will allow us to study the stability of light-quark systems within the
potential model.

After the present manuscript has been submitted for publication, a
recent preprint \cite{Alb06} refers to our procedure of including only
states of sufficient lifetimes and asserts that ``this procedure might
be incorrect, since the evaluated correlator has to be compared with
the lattice ones, which do have a free gas (infinite temperature)
limit."

This statement in Ref. \cite{Alb06} with regard to our work might be
incorrect.  Firstly, the evaluated lattice gauge correlators of
Ref. \cite{Dat03} to be compared were carried out in finite
temperatures with $T \sim (1-3)T_c$ and $T<<T_\infty$ ($T_\infty
\to\infty)$.  This temperature $T$ is in the non-perturbative region
and is far from the infinite temperature perturbative QCD limit.  The
spectrum of a free gas pQCD limit at infinite temperature is not
relevant to the finite temperature lattice gauge correlators being
compared at hand.  Secondly, the presence of a continuum spectrum in
the infinite temperature limit is in fact consistent with our
procedure of including states of sufficient lifetimes at that infinite
temperature limit.

The arguments to support our procedure have been presented already in
the manuscript.  We repeat the main points again below to rebut the
statement of \cite{Alb06}.

As discussed in Eq. (1), the meson correlator describes the
probability amplitude for creating a composite meson at time 0 and
subsequently destroying the composite meson at time $\tau$.  The
operation of destroying the meson at time $\tau$ can be considered as
an operation of measurement (or an operation of detection) of the
meson at time $\tau$.  From the discussions in Eqs. (28) and (29),
$\tau$ is less than and up to $\beta/2$ or $1/2T$, where $T$ is the
temperature, and the measurement operation takes place within
$\tau<1/2T$.  At the temperature $T$, in order to be detected by the
correlator measurement at $\tau=1/2T$, the composite meson state needs
to have a lifetime $\tau_{\rm life}$ greater than $1/2T$, which has been
conveniently labeled as $\tau_{\rm min ~life}(T)=1/2T$ in Eq.\
(\ref{minlife}), the minimum meson state lifetime (for correlator
detection) at temperature $T$.

From this analysis, the composite meson states that can be detected by
the correlator measurement will include meson states of shorter and
shorter lifetimes as temperature increases, and the correlator
spectrum above the bound states region will depend on the temperature.
In the infinite temperature limit $T= T_\infty\to \infty $, the
minimum lifetime $\tau_{\rm min~life}(T_\infty)$, which is equal to
$1/2T_\infty $, approaches zero.  The correlator measurement includes
states that live for a very short composite object lifetime $\tau_{\rm
life}> \tau_{\rm min~life}(T_\infty)\sim 0 $.  This condition can be
satisfied for weakly interacting free gas continuum states. The
correlator allows a free gas continuum spectrum in the infinite
temperature limit.  The presence of a continuum spectrum in the
infinite temperature limit is consistent with our procedure of
including states of sufficient lifetimes.

We turn our attention now to the finite-temperature lattice gauge
calculations in Ref. [24] with $T\sim (1-3) T_c$ and $T<<T_\infty$. Our
afore-mentioned comparison of the meson lifetime and the minimum
lifetime for the correlator measurement indicates that only composite
states that have lifetimes $\tau_{\rm life}$ greater than $\tau_{\rm
min~life}(T)=1/2T$ can survive and be detected by the correlator
measurement at $T$.  Free gas continuum states with very short
composite object lifetimes $\tau_{\rm min~ life}(T_\infty)=
1/2T_\infty$, much smaller than $\tau_{\rm min~ life}(T)= 1/2T$,
cannot survive up to $\tau_{\rm min~ life}(T)=1/2T$ and will not be
detected by the correlator measurement at $T$.  Because of this
limitation, the meson spectrum obtained in finite-temperature
correlator calculations at $T$ should be different from the
correlators in the infinite temperature limit and should only contain
bound states, resonance states, and Gamow states of sufficient
lifetimes greater than $1/2T$.

Based on the above rebuttal, there might be no basis for the authors
in Ref. \cite{Alb06} to make the statement mentioned above.

In conclusion, we have found that the potential model of Ref.\
\cite{Won05} is consistent with both spectral function analyses and
current-current correlator analyses.  The potential model of
\cite{Won05} can be a useful tool to complement lattice gauge
calculations for the study of heavy quarkonia at high temperatures.

\vspace*{0.3cm} This research was supported in part by the Division of
Nuclear Physics, U.S. Department of Energy, under Contract No.
DE-AC05-00OR22725, managed by UT-Battle, LLC and by the National
Science Foundation under contract NSF-Phy-0244786 at the University of
Tennessee and Contract No.\ NSF-PHY-0244819 at the University of
Tennessee Space Institute.


\begin{thebibliography}{99}
\bibitem{footnote} A Gamow state as well as a resonance state is a
  state with a finite lifetime.  For the case of an interacting
  two-body system, a resonance state is further characterized by a
  phase shift of $(2n+1)\pi/2$ where $n$ is an integer, and $\partial
  \delta(\epsilon)/\partial \epsilon >0$.
\end{thebibliography}
\end{document}